\documentclass[doublecol]{epl2}
\usepackage{epsf}
\usepackage{amsmath}
\usepackage{amssymb}

\title{Migration of semiflexible polymers in microcapillary flow}
\shorttitle{Title} %Insert here a short version of the title if it exceeds 70 characters

\author{Raghunath Chelakkot\inst{1} \and Roland G. Winkler\inst{1}
  \and Gerhard Gompper\inst{1}}
\shortauthor{Chelakkot \etal}
\shorttitle{Semiflexible polymers in microchannel flow}

\institute{
  \inst{1}Institut f\"ur Festk\"orperforschung, Forschungszentrum J\"ulich,
  52425 J\"ulich, Germany\\
}

\abstract{
The non-equilibrium structural and dynamical properties of a semiflexible
polymer confined in a cylindrical microchannel and exposed to a
Poiseuille flow is studied by mesoscale hydrodynamic simulations.
For a polymer with a length half of its persistence length, large
variations in orientation and conformations
are found as a function of radial distance and flow strength.
In particular, the polymer exhibits U-shaped conformations near the channel
center. Hydrodynamic interactions lead to strong
cross-streamline migration. Outward migration is governed by the polymer orientation
and the corresponding anisotropy in its diffusivity.
Strong tumbling motion is observed, with a tumbling time which
exhibits the same dependence on Peclet number as a polymer in shear flow.
}

\begin{document}

\maketitle

\section{Introduction}
The flow properties of soft matter under confinement have been
studied intensively in recent years, stimulated by the desire to
unravel the behavior of these materials in nano- and microfluidic
devices. In such systems, confinement effects, a spatially varying
shear rate, and hydrodynamic interactions due to the presence of
impenetrable walls, lead to phenomena not observed in bulk
systems. Often these three effects are mutually competing, and
emergent properties are a result of their combined action.

{\em Flexible polymers} exhibit interesting conformational,
dynamical, and flow properties in the presence of walls and in
channels \cite{jend:03,jend:04,tege:04,bald:07,stei:06,cann:08}.
In particular, cross-streamline migration away from the wall has
been observed
\cite{agar:94,jend:04,usta:06,ma:05,khar:06,stei:06,usta:07,kohl:09,cann:08},
which crucially depends on hydrodynamic lift force exerted by the
wall~\cite{jend:04,ma:05,send:08}. In addition, in channel flow
flexible polymers migrate away from the channel center, resulting
in a decrease in concentration at the centerline
\cite{jend:04,usta:06,cann:08}. This originates from the large
flow-induced polymer stretching and alignment in the high
shear-rate region in the vicinity of a surface, which implies a
smaller lateral diffusion than that of  the coiled conformations
in the channel center. Similarly, migration has been found for
{\em rods} in Poiseuille flow, which has been attributed to their
anisotropic mobility (parallel and perpendicular to the rod axis);
migration towards the surface has been observed when wall
hydrodynamic interactions are neglected
\cite{nits:97,schi:97,sain:06}.

{\em Semiflexible polymers}, such as actin-filaments or short
fragments of DNA, are neither rodlike nor are they able to undergo
large conformational changes. This poses questions as: What is the
influence of hydrodynamics on the polymer distribution? Does
stiffness influence migration? How are the conformations affected
by the flow? Experiments with actin filaments in microchannels
provide evidence for a strong influence of flow on semiflexible
polymers \cite{stei:08}; theory and simulations have demonstrated
the presence of a hydrodynamic lift force in shear flow
\cite{send:08}.

The proper account of hydrodynamic interactions (HI) is essential
in simulation studies of fluid flows in channels as is emphasized
by the appearance of cross-streamline migration. Recently
developed mesoscale simulation techniques, such as Lattice
Boltzmann \cite{ahlr:99,usta:07}, Brownian dynamics with a
hydrodynamic tensor \cite{jend:03}, and multiparticle-collision
dynamics \cite{kapr:08,gomp:09}, are well suited to study
hydrodynamic flows in microchannels and are able to bridge the
length- and time-scale gap between the solvent and solute degrees
of freedom.

\begin{figure}
\begin{center}
\includegraphics*[width=75mm,clip]{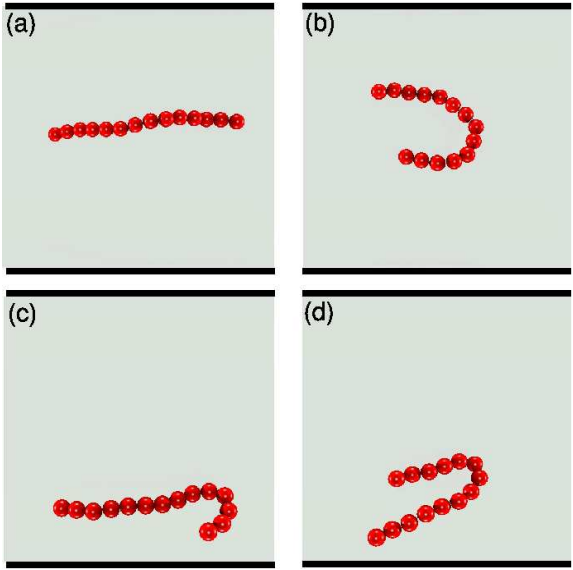}
\caption{Polymer conformations for the Peclet number $Pe =360$.
(a) Aligned and (b) U-shaped conformations in the center of the channel.
(c), (d) Transient hairpin-like conformations outside the center.
The flow direction is from left to right. The channel diameter is indicated by
the black lines \cite{anim}. }
\label{fig:conformations}
\end {center}
\end{figure}

In this letter, we employ multiparticle-collision dynamics (MPC)
to study the properties of a confined semiflexible polymer exposed
to a microchannel flow. We systematically investigate the
influence of the flow rate on the polymer structure,
conformations, dynamics, and distribution functions across the
channel, and provide new insight into the polymer migration
process. Some characteristic polymer conformations in flow are
displayed in fig.~\ref{fig:conformations}. In addition, we discuss
the polymer tumbling dynamics, a property which has been analyzed
previously for flexible molecules in both shear and microchannel
flows \cite{wink:06_1,cann:08}.

\section{Model and simulation method }

We adopt a hybrid simulation approach to study the properties of
semiflexible polymers in flow, where molecular dynamics
simulations (MD) for the polymer are combined with MPC for the
solvent \cite{male:99,ihle:01}. MPC is a particle-based simulation
method and proceeds in two steps. In the streaming step, the
solvent particles of mass $m$ move ballistically for a time $h$.
In the collision step, particles are sorted into the cells of a
cubic lattice of lattice constant $a$ and their relative
velocities, with respect to the center-of-mass velocity of each
cell, are rotated around a random axis by an angle $\alpha$
\cite{kapr:08,gomp:09,ripo:04}. The fluid is confined in a
cylindrical channel with periodic boundary conditions along the
channel axis. No-slip boundary conditions are imposed on the
channel walls by the bounce-back rule and virtual wall particles
\cite{gomp:09,lamu:01}, and flow is induced  by a gravitational
force ($mg$) acting on every fluid particle.

The linear polymer is comprised of $N$ point-like monomers of mass
$M$, which are connected by linear springs with an equilibrium
bond length $b$. Excluded-volume interactions are taken into
account by the truncated $12-6$ Lennard-Jones potential, with
$\sigma$ characterizing the bead size and the energy parameter
$\epsilon$~\cite{muss:05}. The bending potential
\begin{align} \label{v_stiff}
U_b = \frac{\kappa_b}{2} \sum_{i=1}^{N-1} \left({\bm R}_{i+1}-{\bm R}_i \right)^2
\end{align}
is added to account for stiffness~\cite{gomp:09}, where the ${\bm
R}_i$s are bond vectors and the bending rigidity $\kappa_b=L_p k_B
T/b^3$ is related to the persistence length $L_p$ for the
semiflexible polymer. Since we consider pressure-driven flows, no
gravitational force acts on the polymer.

The interaction of a polymer with the solvent is realized by
inclusion of its monomers in the MPC collision
step~\cite{male:00_1}. Between two MPC steps, several MD steps are
performed to update the positions and velocities of the monomers.
Extensive studies of polymer dynamics confirm the validity of this
procedure~\cite{gomp:09,webs:05,muss:05,male:00_1}.

An advantage of the MPC approach is that HI can easily be switched
off, without altering the monomer diffusion significantly
\cite{kiku:03,ripo:07}. In this case, denoted Brownian MPC, each
monomer independently performs a stochastic collision with a
phantom particle with a momentum taken from the Maxwell-Boltzmann
distribution with variance $m \left\langle N_c \right\rangle
k_BT$, where $\left\langle N_c \right\rangle$ is the average
number of solvent particles per collision cell
\cite{gomp:09,ripo:07}.

We employ the parameters $\alpha =130 ^\circ$, $h=0.1 \tau$, with
$\tau=\sqrt{ma^2/k_{B}T}$ ($k_B$ is Boltzmann's constant and $T$
is temperature), $\left\langle N_c \right\rangle =10$, $M=m
\left\langle N_c \right\rangle, b=\sigma =a$, the fluid mass
density $\varrho = \left\langle N_c \right\rangle m/a^3$,
$k_BT/\epsilon =1$, and the time step in MD simulation $h_{MD} =
5\times10^{-3} \tau$. A polymer with $N =14$ monomers is placed in
a cylindrical channel of radius $R = 8.5 a$. With the length
$L_r=(N-1)a=13a$, it does not interact with the wall when its
center of mass is near the channel center. The persistence length
is set to $L_p = 2 Na \approx 2L_r$. Hence, the polymer shows
rodlike behavior at equilibrium. The channel length is $28 a$.
Simulations of the pure solvent system yield velocity profiles
which agree with the solution of Stokes' equation for this
geometry. Averages and probability distributions are calculated in
the stationary state for various independent initial conditions.
To maintain a constant temperature, the velocities are scaled in
every collision step and independently in each collision cell
\cite{huan:10}.

The strength of the applied pressure field is characterized by the
Peclet number, $Pe = {\dot \gamma} \tau_R$, which is here equal to
the Weissenberg number,  where ${\dot \gamma} = g \varrho R /(2
\eta)$ is the shear rate at the cylinder wall and $\tau_R$ the
polymer relaxation time. The Reynolds number $Re = \varrho R v_m/
\eta = \varrho R^2 \dot \gamma/(2 \eta)$, where $v_m$ is the
maximum fluid velocity, depends linearly on the shear rate. For
the above MPC parameters, the viscosity \cite{ripo:04} is such
that $Re < 1$ for all considered $\dot \gamma$. Equilibrium
simulations for a system with periodic boundary conditions yield
the end-to-end vector relaxation time $\tau_R \approx 3200 \tau$.
This value agrees within approximately $20\%$ with the relaxation
time obtained theoretically for a semiflexible polymer with the
same ratio $L_r/L_p$ \cite{wink:06_1}. The relaxation time of the
Brownian MPC simulation is $\tau_R \approx 8200 \tau$.

\begin{figure}
\begin{center}
\includegraphics*[width=70mm,clip]{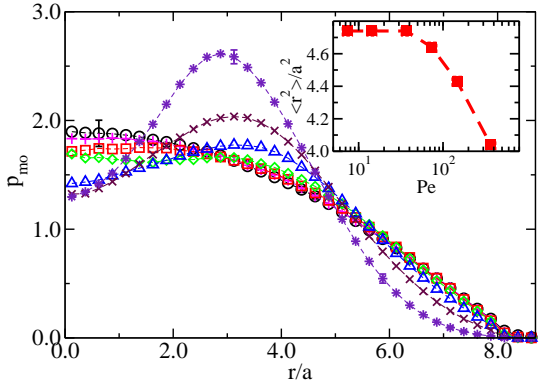}
\caption{
Radial monomer distributions $P_{mo}$ for the Peclet
numbers
$Pe =0$ ($\circ$), $7$ ($+$), $15$ ($\square$), $40$ ($\diamond$),
$70$ ($\triangle$), $150$ ($\times$), and $360$ ($\star$).
Representative error bars are included.
Inset: Widths of the distributions as function of the Peclet number.}
\label{fig1}
\end {center}
\end{figure}

\begin{figure}
\begin{center}
\includegraphics*[width=68mm]{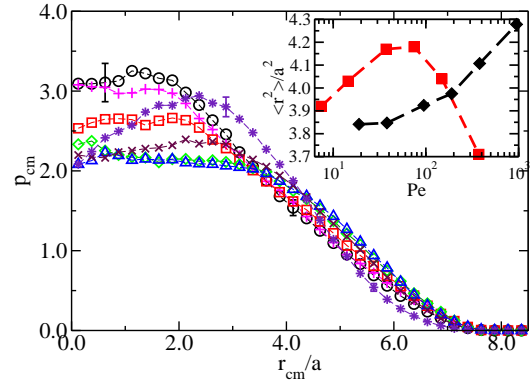}
\caption{Radial center-of-mass distribution functions $P_{cm}$ for the Peclet
numbers $Pe =0$ ($\circ$), $7$ ($+$), $15$ ($\square$), $40$ ($\diamond$),
$70$ ($\triangle$), $150$ ($\times$), and $360$ ($\star$).
Inset: Widths of the distributions as function of the Peclet number with
($\blacksquare$) and without ($\blacklozenge$) HI.}
\label{fig2}
\end {center}
\end{figure}

\section{Radial concentration distributions}

The radial monomer density $P_{mo}$ is displayed in fig.~\ref{fig1}.
It is normalized such that
\begin{align}
\int_0^{R/a} r P_{mo}(r/a) \ dr/a^2=1 .
\label{eqn2}
\end{align}
Without flow, there is a depletion zone close to the channel wall,
which extends approximately one radius of gyration into the
channel. With increasing Peclet number, the density in the center
of the channel decrease, whereas the density near the wall is
unchanged. For even larger $Pe$, we observe a migration of the
molecule away from the surface. At the same time, the density in
the channel center decreases further and a maximum develops at a
finite distance from the center. The maximum is shifted to smaller
radial distances with increasing $Pe$. Two effects contribute to
the formation of the maximum. On the one hand there is
cross-streamline migration of the semiflexible polymer due to
hydrodynamic interactions. On the other hand, the flow field
causes an alignment of the molecule (see discussion below), which
increases the local density at radial positions of strong
flow-induced alignment. The width of the distribution function
\begin{align}
 \langle r^2 \rangle = \int_0^{R/a} r^3 P_{mo}(r/a) \ dr/a^2 ,
\end{align}
shown in the inset of fig.~\ref{fig1}, decreases at Peclet numbers
$Pe \gtrsim 50$, reflecting the inward migration of the polymer.

Figure~\ref{fig2} displays radial center-of-mass distribution
functions $P_{cm}(r_{cm})$ for various Peclet numbers. $P_{cm}$
displays the same qualitative features as the monomer distribution
$P_{mo}$, reflecting the same physical mechanism.  For $Pe
\lesssim 100$, the polymer exhibits a migration away from the
channel center, which is evident from the decrease of $P_{cm}$ in
the central part of the channel, in qualitative agreement with
experiments \cite{stei:08}, and its increase near the channel
wall. This is supported by the broadening  of the width of the
distribution with increasing $Pe$ shown in the inset of
fig.~\ref{fig2}. For $Pe
>100$, the width of the distribution starts to decrease with
increasing $Pe$, indicating the presence of a wall-induced lift
force. As a consequence, the center-of-mass distribution shows a
clear off-centered peak for $Pe \gtrsim {100}$.

It is interesting to compare these results with those of
simulations without HI. We observe an off-center peak in the
monomer density distribution (compare fig.~\ref{fig1}) also
without HI, but the difference between the density in center and
at the peak position is smaller. However, the width $\langle r^2
\rangle$ of $P_{mo}$ does not decrease with increasing $Pe$ in the
absence of HI.  In contrast to the monomer distribution functions,
the distinct off-center peak of the center-of-mass distribution
(fig.~\ref{fig2}) is not present in systems without HI. Here, the
width of $P_{cm}$ increases monotonically with the Peclet number.
Hence, without HI, there is an enhanced outward migration due to
the lack of the wall-induced lift force. At a first glance,
migration without HI seems surprising. However, this is related to
the suppression of steric polymer-wall interactions with
increasing $Pe$ due to flow-induced polymer alignment (cf.
following section). Hydrodynamic interactions enhance the
migration effect, as is seen in fig.~\ref{fig2}. Thus, two effects
contribute to outward migration: alignment of the semiflexible
polymer and intramolecular HI. For the semiflexible polymer, the
latter dominates. As discussed in ref.~\cite{cann:08}, HI are less
relevant for strongly confined flexible polymers.

\begin{figure}
\begin{center}
\includegraphics*[width=70mm,clip]{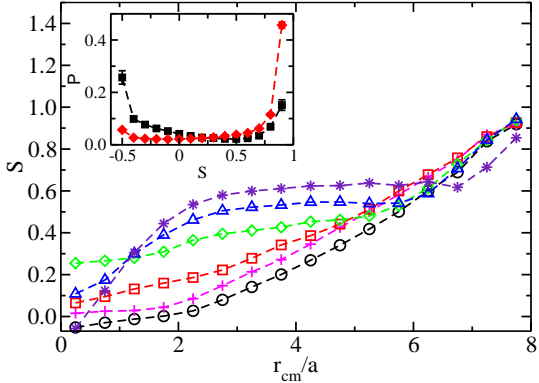}
\caption{Orientational order parameters $S$ as a function of the radial
polymer center-of-mass  position $r_{cm}$ for the Peclet
numbers  $Pe =0$ ($\circ$), $7$ ($+$), $15$ ($\square$), $40$ ($\diamond$),
$70$ ($\triangle$), and $360$ ($\star$). Inset: Probability distribution
$P$ of the orientational order
$S =(3 \cos^2 \theta -1)/2$ for the Peclet number $Pe=360$ and the
radial distance $r_{cm} = 0.5 a$ ($\blacksquare$) and
$r_{cm} = 4.5 a$ ($\blacklozenge$)}
\label{fig3}
\end {center}
\end{figure}

\begin{figure}[t]
\begin{center}
\includegraphics*[width=70mm,clip]{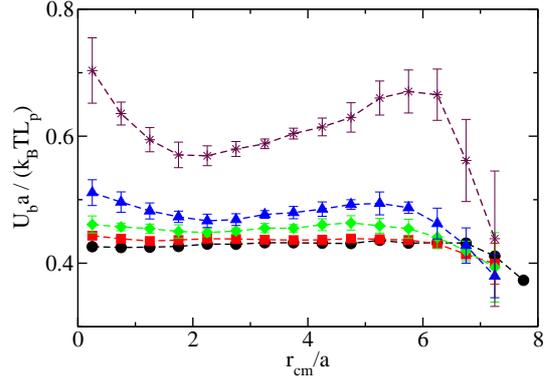}
\caption{Scaled polymer bending energies $U_b$ as a function of radial center-of-mass
position for the Peclet numbers $Pe =0$ ($\bullet$), $15$ ($\blacksquare$),
$40$ ($\blacklozenge$),
$70$ ($\blacktriangle$), and $360$ ($\bigstar$).
The increase in bending energy near the center is an
indication of U-shaped conformations.}
\label{fig5}
\end {center}
\end{figure}

The comparison with confined flexible chains \cite{cann:08}
reveals a stronger inward migration for flexible polymers of
comparable radius of gyration (and thus larger contour length), as
has already been pointed out in ref.~\cite{sain:06}. The reason is
the stretching of the flexible polymer along the flow direction,
which results in a significant longer object than the rodlike
molecule and hence a more pronounced migration. Our studies on the
stiffness dependence of migration for polymers of equal length
show a stronger migration for stiffer molecules.

\section{Polymer conformations and alignment}

In microchannel flow with no-slip boundary conditions, the shear
rate varies linearly across the channel. Hence, the force
experienced by a polymer depends upon its radial position, and its
conformations and alignment are expected to vary as a function of
the radial center-of-mass position. Studies of flexible
polymers~\cite{cann:08} reveal large orientational changes by the
imposed flow, which is important for cross-streamline migration
\cite{send:08}.

Figure~\ref{fig3} displays the orientational order parameter
\begin{equation}
S(r_{cm},\theta)=\frac{1}{2}\langle 3 \cos^{2}\theta - 1 \rangle ,
\end{equation}
where $\theta$ is the angle between the polymer end-to-end vector
and the flow direction, as a function of its center-of-mass radial
position. For $Pe \lesssim 10$, the polymer orientation is
isotropic in the channel center. For large distances from the
center, confinement causes a preferred orientation along the
channel axis; the polymer is almost parallel aligned near the
channel wall. At $Pe \gtrsim 50$, flow induces a strong alignment
even at intermediate $r_{cm}$. Most importantly, at small radial
distances, $S$ decreases again for $Pe \gtrsim 100$, and becomes
even negative for very large Peclet numbers, which we attribute to
the presence of U-shaped conformations (compare
fig.~\ref{fig:conformations}a) -- as observed experimentally
\cite{stei:08}. The probability distributions $P(S)$ for $r_{cm}
\approx 0.5 a$ and $r_{cm} \approx 4.5 a$ reveal the differences
in alignment. At small $r_{cm}$, the probabilities of U-shaped
($S=-0.5$) and aligned straight ($S=1$) conformations are much
larger than those of all other orientations. In contrast, for
larger $r_{cm}$, the probability is largest for almost parallel
aligned polymers. We like to emphasize that the polymer without HI
exhibits very similar orientational order parameters, but only at
significantly larger Peclet numbers.

The large probability $P(S=-0.5)$ (fig.~\ref{fig3}) indicates that
such conformations are rather stable in the channel center. The
snapshot of fig.~\ref{fig:conformations}b shows that despite the
high bending stiffness the polymer adopts  U-shaped conformations.
Such conformations can be quantitatively analyzed by calculating
the average bending energy $U_{b}$. Figure~\ref{fig5} displays
$U_b$ as  function of the radial center-of-mass position and
various Peclet numbers. For $Pe \lesssim 100$ the bending energy
is nearly uniform across the channel cross section and decreases
near the wall, due to wall induced alignment. Its magnitude is
close to the thermal average $U_b = (L_r/a-1)k_B T$ of the nearly
harmonic bending potential, as expected. An increase in $Pe$
results in an increase in its absolute value. For $Pe \gtrsim
100$, $U_b$ is distinctly higher at the channel center due to the
formation of a large number of bend conformations. For polymer
center-of-mass positions slightly out of the center, $U_b$
decreases, since the proportion of bend conformations is reduced.
With increasing radial center-of-mass position, the local shear
rate increases and a polymer assumes transient hairpin
conformations, as shown in figs.~\ref{fig:conformations}c,d. The
average over individual configurations provides a high value for
$U_b$ for such $r_{cm}$. Adjacent to the channel wall the bending
energy decreases again. Here, the wall interactions lead to a
strong alignment with the flow without any bending.

\begin{figure}
\begin{center}
\includegraphics*[width=70mm,clip]{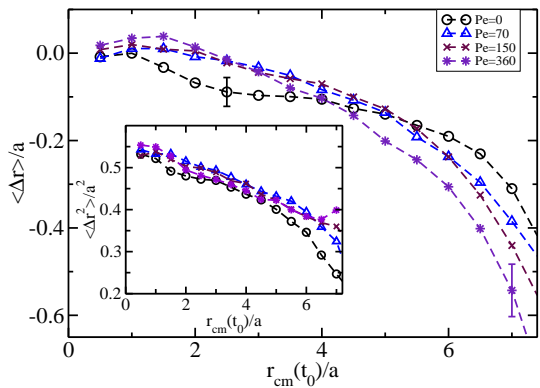}
\caption{Mean radial polymer center-of-mass displacements
$\left\langle \Delta r \right\rangle = \left\langle [{\bm r}_{cm}(t_0+\Delta t)
-{\bm r}_{cm}(t_0)]{\bm e}_{cm}(t_0) \right\rangle$ as function of its radial center-of-mass position
$r_{cm}(t_0)$ for the various
indicated Peclet numbers. Inset:  Radial polymer center-of-mass mean square
displacements $\left\langle \Delta r^2\right\rangle = \left\langle
({\bm r}_{cm}(t_0+\Delta t){\bm e}_{cm}(t_0) -\left\langle {\bm r}_{cm}(t_0+\Delta t) {\bm e}_{cm}(t_0) \right\rangle
)^2\right\rangle$.}
\label{fig6}
\end {center}
\end{figure}

\section{Lateral displacement}

The most striking impact of flow on the semiflexible polymer is
the strong dependence of its orientation on $Pe$ and the radial
distance. This aspect provides the key to understand the
appearance of the off-center maximum in the center-of-mass
distribution function.

As is well known, a rod in solution exhibits a larger diffusion
coefficient parallel to its axis than perpendicular to it. In the
absence of flow, the semiflexible polymer behaves very similar to
a stiff rod, we therefore
%For a rodlike object, the diffusion coefficients
%parallel $D_{\|}$ and perpendicular $D_{\perp}$ to the rod axis
%are different. Averaging over all orientations yields the
%diffusion coefficient for an isotropic configuration:
%\begin{equation}
%D_{iso}=\frac{k_BT\ln(L_r/b)}{3 \pi \eta L_r}= \frac{4}{3}D_{\perp}
%\label{eq:diff}
%\end{equation}
%in the limit of an infinitely long rod, where $L_r$ and $b$ are
%its length and thickness, respectively,  and $\eta$ is the
%viscosity of the fluid. Hence, the isotropic diffusion coefficient
%is larger than the perpendicular one.
expect that the observed polymer orientational differences across
the channel will lead to differences in the lateral dynamic
translational behavior.

In order to characterize this behavior, we present in
fig.~\ref{fig6} the radial center-of-mass displacement $\langle
\Delta r \rangle=
  \langle [{\bm r}_{cm}(t_0+\Delta t) - {\bm r}_{cm}(t_0)]
                         {\bm e}_{cm}(t_0)) \rangle$, ${\bm e}_{cm} = {\bm
r}_{cm}/|{\bm r}_{cm}|$, with respect to the initial radial
position ${\bm r}_{cm}(t_0)$ of the polymer within a time $\Delta
t$, which is chosen such that the center-of-mass displacement is
less than a monomer size $a$ and the mean square displacement is
linear in time. Figure~\ref{fig6} shows a drift away from the wall
for all Peclet numbers and an outward drift for the largest $Pe$
in the channel center, in agreement with the center-of-mass
distribution functions of fig.~\ref{fig2}.  At $Pe=0$, the drift
is governed by steric rod-wall interactions for $r_{cm}\gtrsim
1.5$, which is reflected by the decrease of $\Delta r$ for that
range of $r_{cm}$.  An increase of the flow rate reduces the
steric effect by aligning the polymer, however, the wall lift
force now leads to an increase of the magnitude of the drift for
$r_{cm}/a \gtrsim 5$.

The mean square displacement $\left\langle \Delta r^2\right\rangle
= \left\langle ({\bm r}_{cm}(t_0+\Delta t) {\bm e}_{cm}(t_0)
-\left\langle {\bm r}_{cm}(t_0+\Delta t) {\bm
e}_{cm}(t_0)\right\rangle )^2\right\rangle$, shown in the inset of
fig.~\ref{fig6}, displays only a slight dependence on drift for
$r_{cm}/a \lesssim 5$; its decrease with increasing $r_{cm}$
reflects the dependence of the mean square displacement on radial
position. Without flow, the polymer orientation is isotropic in
the central part of the channel, and it is essentially parallel to
the channel axis near the wall. Hence, we see the difference
between isotropic and perpendicular diffusive motion originating
from intramolecular hydrodynamic interactions. For larger radii,
steric wall interactions lead to a decrease of the mean square
displacement for low Peclet numbers, whereas lift forces increase
$\left\langle \Delta r^2\right\rangle$ for larger $Pe$ values.

Non-hydrodynamic simulations yield constant mean square
displacements for $r_{cm}/a \lesssim 5$ for the considered range
of Peclet numbers. At larger $r_{cm}$, steric wall interactions
lead to a decrease of $\left\langle \Delta r^2\right\rangle$,
similar to that for $Pe=0$ in fig.~\ref{fig6}.

Hence, we conclude that the center-of-mass distribution function
is determined by a competition of the radial outward diffusion of
a flow-aligned rod and inward migration caused by a wall-induced
lift force \cite{send:08}.

\section{Polymer tumbling}

\begin{figure}
\begin{center}
\includegraphics*[width=70mm,clip]{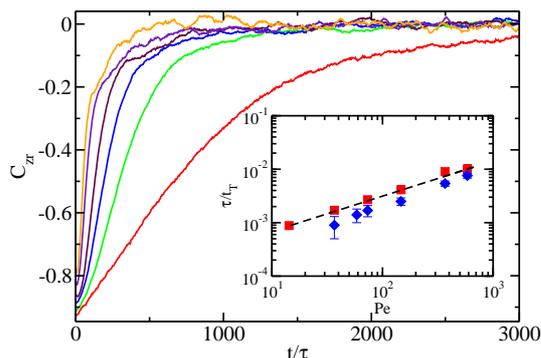}
\caption{Cross-correlation functions $C_{zr}$ for the
Peclet numbers $Pe = 15, 40, 70, 150, 360$, and $570$ (bottom to top).
Inset: Tumbling times $t_T$ calculated from the cross-correlation
function $C_{zr}$ ($\blacksquare$) and the probability distribution of time
intervals between radial orientations of the end-to-end vector ($\blacklozenge$). }
\label{fig7}
\end {center}
\end{figure}

Flexible polymers in shear and Poiseuille flow undergo large
conformational changes by tumbling motion, i.e., the polymers
exhibit a cyclic stretching and collapse dynamics
\cite{wink:06_1,teix:05}. Visual examination of the dynamics of
the semiflexible polymer also indicates tumbling motion. This
cyclic motion is characterized by the tumbling time $t_T$.
Analytical calculations for a planar shear flow have shown that
$t_T \sim {\dot \gamma}^{-2/3}$ for flexible and rodlike polymers,
where $\dot \gamma$ is the shear rate \cite{wink:06_1}. In order
to compute the tumbling time for different $Pe$, we determine the
cross-correlation function between the mean extensions $\delta z =
z - \left\langle z \right\rangle$ and $\delta r = r - \left\langle
r \right\rangle$ in flow and radial directions, respectively,
which is defined as
\begin{equation}
C_{zr}(t) = \langle \delta z(t_0) \delta r(t_0+t) \rangle/
\sqrt{\langle \delta z^2(t_0)\rangle \langle \delta r^{^2}(t_0)\rangle}.
\end{equation}
The time dependence of the correlation function is presented in
fig.~\ref{fig7}. Microchannel flows are  more complex in nature
compared to planar shear flows. In planar shear flow, the polymer
exhibits essentially a two-dimensional behavior, whereas the
channel flow is rotational symmetric with respect to the channel
axis and more complex conformational transitions appear.
Therefore, $C_{zr}$ is qualitatively different from to a similar
correlation function in planar shear flow for flexible
polymers~\cite{teix:05}. In the latter case, at $Pe>1$, the
correlation function is asymmetric and displays a minimum and a
maximum, whereas we obtain a time-symmetric negative correlation
function with a peak at $t=0$. The width of the correlation
function decreases with increasing flow strength, which is a
measure for the dependence of $t_T$ on $Pe$. Hence, we quantify
the tumbling time by the full width at half maximum of $C_{zr}$
for each value of $Pe$. As shown in the inset of fig.~\ref{fig7},
the tumbling time decreases as $t_T \sim Pe^{-0.66}$, i.e., the
exponent is very close to the analytically predicted exponent for
planar shear flow \cite{wink:06_1}. In order to verify the result,
a tumbling time is also extracted from the distribution function
of the time intervals between consecutive crossings of the
end-to-end vector with the plane perpendicular to the flow
direction~\cite{cann:08}. This distribution decays exponentially
($\exp(-t/t'_T)$) for sufficiently large times. The tumbling time
$t'_T$ exhibits the same dependence on the Peclet number as $t_T$
obtained from $C_{zr}$; quantitatively it is smaller by
approximately a factor 1.5. Hence, both definitions of tumbling
time are equivalent for the considered system.

\section{Summary and Conclusions}

We have analyzed the flow behavior of an actin-like semiflexible
polymer confined in a cylindrical channel. The Poiseuille flow
induces a strong radial-dependent polymer alignment parallel to
the channel associated with a radially outward migration of its
center of mass at low Peclet numbers \cite{nits:97,schi:97} and an
inward migration at large $Pe$; both effects are due to
hydrodynamic interactions.

This behavior is surprisingly similar to that of a flexible
polymer confined in a channel
\cite{jend:03,jend:04,cann:08,usta:06,ma:05,khar:06,usta:07}. It
is caused by the common (flow induced) anisotropy of the polymers,
which is responsible for the appearance of migration
\cite{send:08}. Flexible polymers, however, exhibit much more
pronounced conformational changes and large deformations, which
depend on the radial center-of-mass position, but there is no
enhanced probability for U-shaped conformations. We also find
different dependencies of the tumbling time on the flow strength
\cite{cann:08}.

Our studies confirm the importance of HI in microchannel flows.
Nevertheless, astonishingly similar distributions and
conformations are obtained in simulations without HI, but they
appear at significantly larger Peclet numbers. However, in this
large $Pe$ regime, hydrodynamics determines the distributions in
the channel. Thus, non-hydrodynamic simulations may provide
reasonable structural and conformational properties of a fully
hydrodynamic system as long as wall lift forces are negligible,
but will fail for large $Pe$ even qualitatively. \\
\vspace*{-0.8cm}

%The alignment of the rods leads to an inhomogeneous radial mean
%square displacement, which is faster in the channel center
%\cite{nits:97,schi:97}, and is a consequence of intramolecular
%hydrodynamic interactions. The competition between of the faster
%outward diffusion and wall hydrodynamics induced inward migration
%leads to appearance of an off-center center-of-mass density
%maximum.

%The semiflexible polymer exhibits tumbling dynamics, where bend
%conformations appear at high Peclet numbers. The calculation of
%the characteristic tumbling time by the extension
%cross-correlation function and the distribution function of time
%intervals yield the same dependence on the Peclet number. The
%shape of the correlation function clearly reflects the
%non-periodicity of the tumbling motion. Moreover, the quantitative
%agreement of the tumbling times underlines the usefulness of the
%correlation function to determine $t_T$.

\acknowledgments We thank Thomas Pfohl (Basel) for stimulating
discussions. The financial support by the Deutsche
Forschungsgemeinschaft within the Priority Program ``Nano- and
Microfluidics" (SPP 1164) is gratefully acknowledged.\\
%\vspace*{-0.7cm}
%\bibliographystyle{eplbib}
%\bibliography{polymer_channel}

\end{document}